\documentclass{aa}  
\usepackage{ulem}
\usepackage{graphicx}
\usepackage{natbib}
\usepackage[colorlinks,citecolor=blue,linkcolor=blue,urlcolor=blue]{hyperref}
\usepackage{color} 
\usepackage{amsmath}
\usepackage{fixltx2e}
\def\ii{{\text{\tiny i}}}
\def\co{{\text{\tiny cut-off}}}
\def\d{{\mathrm{d}}}
\def\PBH{{\mathrm{PBH}}}
\def\lp{\left (}
\def\llp{\left [}
\def\rp{\right )}
\def\rrp{\right ]}

\definecolor{ao(english)}{rgb}{0.0, 0.5, 0.0}

\definecolor{chmagenta}{rgb}{0.54, 0.17, 0.88}

\usepackage{graphicx}
\usepackage{txfonts}

\usepackage{soul}

\begin{document}

   \title{Stochastic gravitational-wave background as a tool to investigate  multi-channel astrophysical and primordial black-hole mergers}
    
   \titlerunning{Constraining model selection of astrophysical and primordial BBHs with the SGWB}

   \author{Simone\,S.\,Bavera
          \inst{1}
          \and
          Gabriele\,Franciolini
          \inst{2}
          \and
          Giulia\,Cusin
          \inst{2}
          \and
          Antonio\,Riotto
          \inst{2}
          \and
          Michael\,Zevin
          \inst{3,4}
          \and 
          Tassos\,Fragos
          \inst{1}
          }
    \authorrunning{Bavera et al.}

   \institute{Departement d’Astronomie, Université de Genève, Chemin Pegasi 51, CH-1290 Versoix, Switzerland\\ \email{simone.bavera@unige.ch}
    \and
    Département de Physique Théorique and Centre for Astroparticle Physics, Université de Genève, 24 quai E. Ansermet, CH-1211 Geneva, Switzerland
    \and
    Kavli Institute for Cosmological Physics, The University of Chicago, 5640 South Ellis Avenue, Chicago, Illinois 60637, USA
    \and
    Enrico Fermi Institute, The University of Chicago, 933 East 56th Street, Chicago, Illinois 60637, USA
             }

   \date{Accepted December 21, 2021}

 
  \abstract{
  The formation of merging binary black holes can occur through multiple astrophysical channels such as, e.g., isolated binary evolution and dynamical formation or, alternatively, have a primordial origin. 
  Increasingly large gravitational-wave catalogs of binary black-hole mergers have allowed for the first model selection studies between different theoretical predictions to constrain some of their model uncertainties and branching ratios.
  In this work, we show how one could add an additional and independent constraint to model selection by using the stochastic gravitational-wave background. 
  In contrast to model selection analyses that have discriminating power only up to the gravitational-wave detector horizons (currently at redshifts $z\lesssim 1$ for LIGO--Virgo), the stochastic gravitational-wave background accounts for the redshift integration of all gravitational-wave signals in the Universe.
  As a working  example, we consider the branching ratio results from a model selection study that includes potential contribution from astrophysical and primordial channels. We renormalize the relative contribution of each channel to the detected event rate to compute the total stochastic gravitational-wave background energy density. The predicted amplitude lies below the current observational upper limits of GWTC-3 by LIGO--Virgo, indicating that the results of the model selection analysis are not ruled out by current background limits. 
  Furthermore, given the set of population models and inferred branching ratios, we find that, even though the predicted background will not be detectable by current generation gravitational-wave detectors, it  will be accessible by third-generation detectors such as the Einstein Telescope and space-based detectors such as LISA.
  }


   \maketitle
%

\section{Introduction}\label{sec:intro}

   Coalescing binary black holes (BBHs) are the primary sources of gravitational-waves (GWs) currently detectable by the LIGO \citep{2015CQGra..32g4001L}, Virgo \citep{2015CQGra..32b4001A} and KAGRA \citep{2013PhRvD..88d3007A} detectors. To date, counting only events with false alarm rate of $< 1\,\mathrm{yr}^{-1}$, the signals of 69 confident BBH mergers have been reported along with 2 binary neutron stars (BNSs), 2 black hole--neutron star (BHNS) systems \citep{2019PhRvX...9c1040A,2021ApJ...915L...5A,2021PhRvX..11b1053A,2021arXiv210801045T,2021arXiv211103606T}.
   
    In addition to events that are individually detectable, the entire population of unresolved and resolved events generates a stochastic gravitational-wave background (SGWB) signal. Other than compact binary coalescences, there are multiple astrophysical and cosmological sources contributing to the SGWB. Possible sources include core-collapse supernovae, magnetars, cosmic strings and GWs produced during inflation \citep[e.g.][]{2011RAA....11..369R,2020arXiv200109663D}. However in the frequency ranges of current GW observatories, the SGWB is thought to be dominated by coalescing compact binaries \citep{2016PhRvL.116m1102A,2018PhRvL.120i1101A} where BBHs mergers are thought to dominate the SGWB over BNS or BHNS systems \citep{2020arXiv200109663D,2021PhRvD.103d3002P,2021arXiv211201119P}. 
   
   The SGWB is characterized by a spectral energy spectrum, $\Omega_\mathrm{GW}(\nu)$, which can be measured by cross-correlating the data streams from multiple detectors \citep{1992PhRvD..46.5250C,1999PhRvD..59j2001A}. Using the data of the first three observing runs (O1, O2 and O3), the LIGO Scientific and Virgo Collaboration (LVC) did not find evidence of the SGWB. Hence, the LVC was able to put an upper limit to the SGWB energy density spectrum of  $\Omega_\mathrm{GW}(\nu = 25\,\mathrm{Hz}) \leq 1.04 \times 10^{-9}$ for a power-law SGWB with a spectral index of $2/3$, consistent with expectations for coalescing compact binary \citep{2021arXiv211103634T}. 
   
   
   Multiple astrophysical evolutionary pathways may contribute to the formation of coalescing BBHs, which are often divided into categories. The isolated binary evolution family of pathways includes binaries evolving through a stable mass transfer (MT) and a common envelope (CE) phase \citep[e.g.][]{1998ApJ...506..780B,2007PhR...442...75K,2014LRR....17....3P,2016Natur.534..512B,2020A&A...635A..97B}, double stable MT (SMT) \citep[e.g.,][]{2017MNRAS.471.4256V,2019MNRAS.490.3740N,2021A&A...647A.153B},  and chemically homogeneous evolution (CHE) \citep[e.g.][]{2009A&A...497..243D,2016MNRAS.458.2634M,2016A&A...588A..50M,2020MNRAS.499.5941D}.  Dynamical formation of BBHs in dense stellar environments may occur in globular clusters (GCs) \citep[e.g.][]{1993Natur.364..423S,2002MNRAS.330..232C,2010MNRAS.407.1946D,2015PhRvL.115e1101R}, nuclear star clusters (NSCs) \citep[e.g.][]{2009ApJ...692..917M,2017ApJ...846..146P,2019MNRAS.486.5008A,2020ApJ...891...47A} or young open star clusters \citep[e.g.][]{2014MNRAS.441.3703Z,2016MNRAS.459.3432M,2017MNRAS.467..524B,2020MNRAS.495.4268K}. Population III stars have also been proposed to lead to merging BBH either in isolation or in stellar cluster \citep[e.g.][]{2001ApJ...551L..27M,2014MNRAS.442.2963K,2017MNRAS.468.5020I,2021MNRAS.501..643L}. 
   Furthermore, alternative proposed channels exists such as the formation of merging BBHs in active galactic nuclei disks \citep[e.g.][]{2012ApJ...757...27A,2014MNRAS.441..900M,2017ApJ...835..165B,2020ApJ...899...26T} and in triple or multiples systems \citep[e.g.][]{2016ApJ...816...65A,2017ApJ...841...77A,2019MNRAS.486.4443F,2021ApJ...907L..19V}. 

    Another well-studied formation channel for producing merging BBHs is through primordial origins (PBHs)~\citep{Zeldovich:1967lct,Hawking:1974rv,Chapline:1975ojl,Carr:1975qj}. PBHs may arise from the collapse of large overdensities in the radiation-dominated early universe~\citep{Ivanov:1994pa,GarciaBellido:1996qt,Ivanov:1997ia,Blinnikov:2016bxu} and could contribute to a sizeable ratio $f_\PBH\equiv \Omega_\PBH/\Omega_\text{\tiny DM}$ of the dark matter energy density in a variety of mass ranges \citep[see][for a recent review about constraints on $f_\PBH$]{2020arXiv200212778C}. The recent discovery of GWs has ignited a new wave of interest on PBHs as it was soon realised that PBHs could produce observable mergers without conflicting with existing bounds on the PBH abundance \citep{Bird:2016dcv,Clesse:2016vqa,Sasaki:2016jop}. This motivated various works on the confrontation of the PBH scenario with the most recent data \citep[see, for example, the recent results of ][]{2016arXiv161008725W,2019PhRvD..99j3531W,2020PhRvD.102l3524H, 2021JCAP...05..039K,Hutsi:2020sol,2021JCAP...05..003D,Deng:2021ezy,Kimura:2021sqz}.
    Current GW data imply an upper bound $f_\PBH \lesssim {\cal O}(10^{-3})$ in the mass range of interest for current GW detectors \citep[see e.g.,][]{Wong:2020yig}.
    The constraining power of GW observations of either resolved mergers or the SGWB will improve significantly with future GW detectors \citep{2021arXiv210613769D,2021arXiv210703379P}.

    All these channels have been shown to successfully lead to the formation of merging BBHs
    and, in most cases, predict a plausible range of merger-rate density approximately consistent with current GW observational constraints. However, accurate rate estimates are often difficult to be made, as they are highly dependent on uncertain and sometimes unconstrained astrophysical processes. The most well-known uncertainties affecting the astrophysical models include initial stellar and binary properties (e.g., binary ratio, initial mass function, mass ratio and initial orbital parameter distributions), stellar evolution physics (e.g., stellar winds of massive stars, core-collapse mechanism, supernova kicks and pulsational pair instability), binary evolution physics (e.g., MT stability and efficiency, and CE efficiency) as well as uncertainties in the star formation rate and metallicity distribution of their environment at high redshifts \citep[ see e.g.,][]{2017ApJ...841...77A,2017ApJ...834...68C,2019MNRAS.482.5012C,2019MNRAS.490.3740N,2020A&A...638A.119G,2021MNRAS.505..663R, 2021arXiv210810885B}. The primordial channel also suffers from large uncertainties on the overall PBH abundance and initial mass distribution, which are mostly unconstrained in the mass range of interest for LVC mergers \citep{2020arXiv200212778C}. Combined, these unconstrained physical processes lead to order-of magnitudes uncertainties in the rates while often have minor effects on the BBH observable distributions \citep[see e.g.,][for a review]{2021arXiv210714239M}. Such large uncertainties translate to the SGWB energy spectrum and also bias relative $\Omega_\mathrm{GW}$ estimates between channels.

   Given the large uncertainty on the modelled BBH rates, comparison between theoretical predictions and GW observations are often done by normalising the theoretical rate to the observed one. Recent attempts of model selection involving multiple formation channels and GWTC-2 events \citep{2021PhRvD.103h3021W,2021ApJ...910..152Z,2021arXiv210503349F} indicate that, given the state-of-the-art formation models considered, multiple formation channels are needed to explain the detected population of BBHs. To date, \citet{2021ApJ...910..152Z} work is the most inclusive analysis which accounts for CE, SMT, CHE, GC and NSC channels. Even though \citet{2021ApJ...910..152Z} find that a mixture of channels is preferred over a single dominant channel, at face value, the analysis shows that isolated BBH formation might dominate the underlying BBH merging population. It is important to note that only the uncertainties of CE efficiency \citep[cf.][]{2021A&A...647A.153B} and isolated BH spin as a proxy for angular momentum transport were explored in that analysis. The consideration of all model uncertainties and the other prominent BBH formation channel are key in obtaining an unbiased and conclusive answer, though the large number of proposed formation models and model uncertainties, combined with the still limited observational sample, make this task currently computationally infeasible. 
   Following an analysis similar to \citet{2021ApJ...910..152Z}, \citet{2021arXiv210503349F} expanded the set of considered models by also including the PBH channel. The Bayesian evidence for PBHs against an astrophysical-only multi-channel model was found decisive, due to the fact that the astrophysical models considered there did not produce significant numbers of high-mass events, such as those observed by the LVC. However, the evidence weakens in the presence of a dominant SMT isolated formation channel, which is more efficient at producing high-mass BBHs.
   
   
   In this work we compute the contribution to the SGWB energy spectrum of astrophysical and primordial BBHs using the results of the model comparison by \citet{2021arXiv210503349F} as a working example. The relative contribution of each channel is directly dictated by the comparison of the models with the GWTC-2 data, and their total contribution normalised against the BBH detection rate of 44 events, including GW190521.\footnote{Following \cite{2021ApJ...913L...7A}, \cite{2021arXiv210503349F} discarded from the  GWTC-2 catalog the events with large false-alarm rates (GW190426, GW190719, GW190909) and two events involving neutron stars (GW170817, GW190425). Also, as the secondary mass ($m_2 \approx 2.6\,\mathrm{M}_\odot$) of GW190814 \citep{2020ApJ...896L..44A} 
  would correspond to either the lowest-mass astrophysical BH or to the highest-mass NS observed to date, 
  challenging current understanding of compact objects, it was assumed that the secondary component of GW190814 is a neutron star and this event was neglected. Notice, however, that its inclusion would not affect our conclusions \citep{2021arXiv210503349F}.} This assumption is complementary to previous studies
    which use phenomenological astrophysical models \citep[e.g.][]{2020ApJ...896L..32C,2021MNRAS.500.1421Z} and arbitrarily fix the relative contribution of primordial to astrophysical BBHs \citep[e.g.][]{2019ApJ...871...97C,2021MNRAS.506.3977M,2021arXiv210702181M}. Moreover, our analysis gives a further model constraint to model selection studies as the SGWB includes the rate contribution across all redshift compared to current GW detectors horizons of $z\lesssim1$ \citep{2020PhRvL.125j1102A}. The results presented in this work are meant to illustrate a methodology which can be applied to and extended with future catalogs of BBH events, putting more stringent constraints on model selection studies. The best fit parameters of  astrophysical and PBH models, with the corresponding  branching ratios,  can change drastically with adjustments in the population models or the formation channels considered. However, accounting for the measurement or upper limits on the SGWB  provides a new additional constraint on the validity of model selection results, which to our best knowledge is considered here for the first time.
   
   The paper is structured as follows. In Section~\ref{sec:methods} we present the main assumption of each considered astrophysical and primordial BBH channel and explain how we estimate the SGWB energy density spectrum. Section~\ref{sec:results} presents the SGWB energy density spectrum of our models compared to current and planned ground- and space-based GW observatories such as the Einstein Telescope (ET) \citep{2010CQGra..27s4002P} and the Laser Interferometer Space Antenna (LISA) \citep{2017arXiv170200786A}. 
   Finally, in Section~\ref{sec:discussion} we discuss how our results depend on model uncertainties and we quantify the effect of neglecting other prominent channels. In Section~\ref{sec:conclusions} we summarise our findings.

\begin{figure*}
\centering
\includegraphics[width=0.49\linewidth]{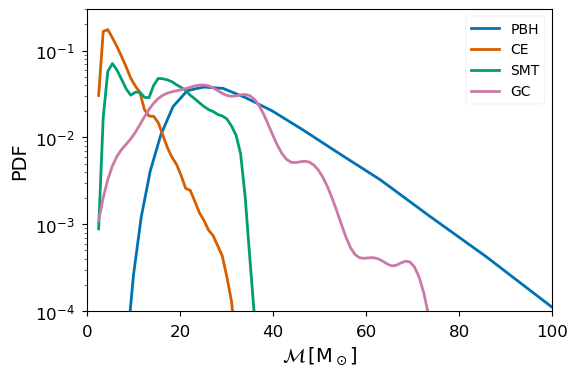}
\includegraphics[width=0.49\linewidth]{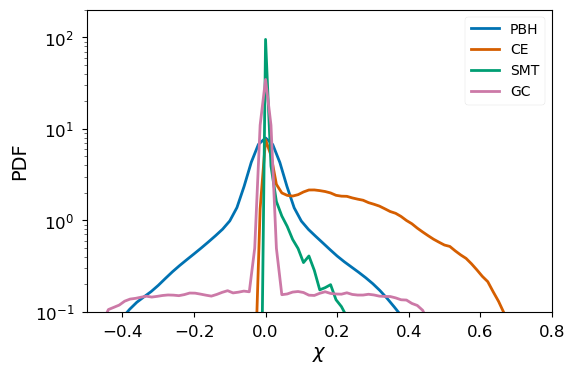}
\caption{Gravitational-wave observables for the intrinsic distribution of merging BBHs in the Universe for different formation channels according to the legends. (\textit{Left}) We show the normalised chirp mass, $\mathcal{M}$, distributions where we can see that the PBH and GC channels leads to more massive BHs, whereas the maximum BH mass of CE and SMT channels is dictated by pulsational pair instability supernovae.
(\textit{Right}) We show the normalised effective spin parameter, $\chi$, distributions where we see that only the CE channel generates a large ratio of positive $\chi$ due to tidal spin up, while both the GC and PBH channels lead to a symmetric distribution of $\chi$ (allowing for negative values) because of isotropically oriented spins in hierarchical mergers (GC) or uncorrelated spin growth (PBHs). 
}
\label{fig:observable_dist}
\end{figure*}

\section{Methods}\label{sec:methods}

    We compute the SGWB energy density spectrum of merging BBHs from astrophysical and primordial origins. For the astrophysical channels we include isolated binary evolution evolving through CE and SMT channels, and dynamical formation in GC. We adopt BBH models of isolated binary evolution by \citet{2021A&A...647A.153B}, a GC model by \citet{2019PhRvD.100d3027R} as released by \citet{2021ApJ...910..152Z} and a PBH model by \citet{2021arXiv210503349F}. The key assumptions of all these models are summarised in Appendix~\ref{app:BBH_models}. More precisely, as favoured by the model comparison with GWTC-2 data in \citet{2021arXiv210503349F} \citep[but also see][]{2021ApJ...910..152Z}, we use models with isolated BHs birth spins of zero and for the CE channel the model with common envelope efficiency $\alpha_\mathrm{CE}=5$. Here, an $\alpha_\mathrm{CE}$ value grater than 1 does not mean that other sources of energy partake in the CE ejection, but more likely points to an inaccurate assumption of core-envelope boundaries in the $\alpha_\mathrm{CE}-\lambda$ parametrization of CE \citep[see, e.g.][for a review]{2013A&ARv..21...59I}. The fact that envelope stripping stops earlier than what currently assumed in population synthesis models has been suggested in multiple recent studies \citep{2019ApJ...883L..45F,2019A&A...628A..19Q,2021A&A...645A..54K,2021A&A...650A.107M}. Finally, the combined and relative detection rate and, hence indirectly, the local merger rate density of these channels are calibrated against the model selection comparison of \citet{2021arXiv210503349F} with GWTC-2 events.
    
    For a graphical visualisation of the intrinsic distributions of the main BBH observables of all considered channels, in Figure~\ref{fig:observable_dist}, we show the underlying distributions of chirp mass $\mathcal{M}$ and effective spin parameter $\chi$. The intrinsic (underlying) BBH distribution is what a GW detector with infinite sensitivity would observe on Earth. Here, the chirp mass is defined as $\mathcal{M}=(m_1m_2)^{3/5}/(m_1+m_2)^{1/5}$ where $m_1$ and $m_2$ are the BH masses and the effective spin parameter $\chi = (m_1 {\bf{a}}_1 + m_2 {\bf{a}}_2)/M \cdot {\bf{\hat{L}}}$ where $\bf{a}_1$ and $\bf{a}_2$ are the BHs dimensionless spins and $\bf{\hat{L}}$ is the orbital angular momentum unit vector. The probability density functions (PDFs) of astrophysical channels are constructed using kernel density estimators (KDE) on the BBH discrete model results, while for the PBH channel the PDFs are semi-analytically determined using Eq.~\eqref{diffaccrate}.
    
    In Figure~\ref{fig:observable_dist}, the maximum $\mathcal{M}$ of the isolated binary evolution channels CE and SMT is dictated by pulsational pair instability supernovae \citep[e.g.,][]{1964ApJS....9..201F,2017ApJ...836..244W,2019ApJ...882...36M} while this is not the case for the GC channel since merger products can be retained in the cluster and merge again.
    PBHs form from the collapse of large overdensities in the early universe and their mass is related to the mass contained in the cosmological horizon at the time of collapse. For this reason, PBHs can form in a much wider range of masses compared to astrophysical BHs and can cover the mass gap \citep[e.g.][]{2021PhRvL.126e1101D}.
    Even though we assume astrophysical BHs are born with zero spin in isolation, tidal interactions in the later phase of close BH-Walf-Rayet systems can tidally spin up second born BHs in the CE and SMT channels \citep[e.g.,][]{2018A&A...616A..28Q,2020A&A...635A..97B}. The spin of the resultant BH is mostly aligned with the orbital angular momentum since BH natal kicks are not typically strong enough to flip the orbits by more than $90^\circ$ \citep[e.g.,][]{2016ApJ...832L...2R,2020arXiv201109570C}. Hence, the minimum effective spin parameter of CE and SMT channels is $\chi \simeq 0$. In contrast, GC channel might lead to negative $\chi$ given the random dynamical assembly of the BBH systems. Since the majority of these systems are the results of first-generation mergers whose components do not proceed through tidal spinup, the effective spin distribution peaks at $\chi \simeq 0$. However, merger products retained in the cluster are imparted spin due to the angular momentum of the inspiraling binary, and thus hierarchical mergers lead to a symmetric distribution of $\chi$ about zero. 
    While PBHs form with negligible spin in the standard scenario, efficient accretion can spin up PBHs along independent directions and lead to large magnitudes for $\chi$ symmetrically distributed around zero, see more details in Appendix~\ref{app:PBHs}. 
    
    The SGWB energy density spectrum, $\Omega_\mathrm{GW}$, for each model as well as the GW detectors' power-law-integrated sensitivity curves are calculated as explained in the following Sections. 

\subsection{SGWB energy density spectrum}\label{sec:SGWB_Omega}

    Under the cosmological assumption of the $\Lambda$CDM model, the fractional energy density content of the Universe today is dominated by matter $\Omega_m \simeq 0.307$ and dark energy $\Omega_\Lambda \simeq 1 - \Omega_m$. These energy density ratios are defined in terms of closure density $\rho_c = 3 H_0^2/(8\pi G)$ with a current Universe relative rate of expansion $H_0 = 67\mathrm{km/s}/\mathrm{Mpc}$ \citep{2016A&A...594A..13P}. 
    
    In comparison, the energy density ratio of the SGWB, $\Omega_\mathrm{GW}$, is small and often expressed as a spectrum, namely as a function of frequency. This is done in order to compare it with GW detectors' power-law-integrated sensitivity curves which are frequency dependent. Here, we consider frequencies of current ad future ground-based detectors such as LIGO, Virgo, KAGRA and ET, which are sensitive to the $[1,10^3]\,\mathrm{Hz}$ band, as well as  space-based detectors like LISA, which are sensitive to $[10^{-4},0.1]\,\mathrm{Hz}$. In such bands, the spectral GW energy density is dominated by merging BBHs \citep{2020arXiv200109663D,2021PhRvD.103d3002P}.

\begin{figure}
\centering
\includegraphics[width=\linewidth]{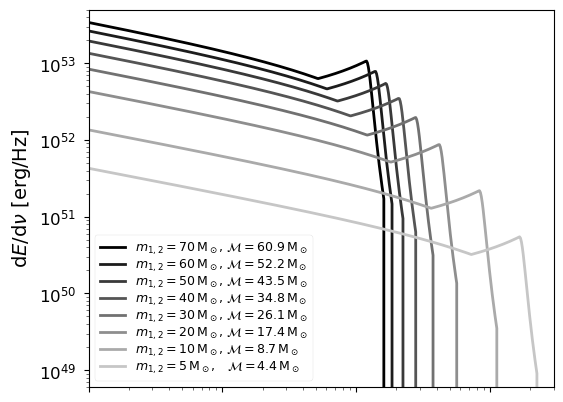}
\includegraphics[width=\linewidth]{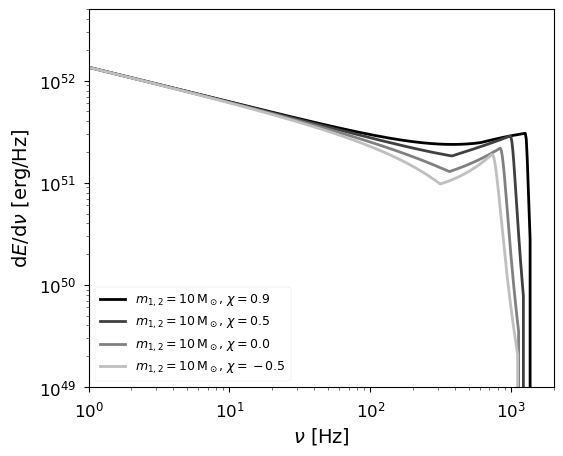}
\caption{Gravitational-wave energy spectrum, $\mathrm{d}E/\mathrm{d}\nu$, as a function of frequency. (\textit{Top}) We show the energy spectrum for different non-spinning equal mass BBHs according to the legend. The more massive the BBH systems is, the more energetic the GW signals and the smaller the merger frequencies. More precisely, the GW energy spectrum scales as $\mathrm{d}E/\mathrm{d}\nu \propto \nu^{-1/3} \mathcal{M}^{5/3}$ for $\nu < \nu_\mathrm{merger}$ and $\mathrm{d}E/\mathrm{d}\nu \propto \nu^{2/3} \mathcal{M}^{5/3}$ for $\nu < \nu_\mathrm{ringdown}$ (cf. Eq.~\ref{eq:dEdnu}), while $\nu_\mathrm{merger} \propto M^{-1}$ where $M=m_1+m_2$ is the total mass (cf. Eq.~\ref{eq:nus}).
(\textit{Bottom}) We show the energy spectrum for a BBH system composed of two BHs of mass $10\, \mathrm{M}_\odot$ and different effective spins according to the legend. Larger $\chi$ values lead to more energetic GW signals as $\mathrm{d}E/\mathrm{d}\nu \propto \nu^{-1/3} \left(1+{\cal O}(\chi)\right)$ for $\nu < \nu_\mathrm{merger}$ and $\mathrm{d}E/\mathrm{d} \nu \propto \nu^{2/3} \left(1+{\cal O}(\chi)\right)$ for $\nu < \nu_\mathrm{ringdown}$ (cf. Eq.~\ref{eq:dEdnu}).
}
\label{fig:dEdnu}
\end{figure}

    The SGWB spectral energy density ratio is defined as \citep[e.g.][]{2011ApJ...739...86Z}
    \begin{equation}
        \Omega_\mathrm{GW}(\nu_\mathrm{obs}) = \frac{\nu_\mathrm{obs}}{c^3 \rho_c} F_\mathrm{\nu}(\nu_\mathrm{obs}) \, ,
        \label{eq:OmegaGW}
    \end{equation}
    where $\nu_\mathrm{obs}$ is the observed GW frequency related to the source frame by $\nu = \nu_\mathrm{obs}(1+z)$, $F_\mathrm{\nu}$ is the GW spectral energy density 
    and $c$ the speed of light. Here, we can express
    \begin{equation}\label{eq:Fnu}
        F_\mathrm{\nu}(\nu_\mathrm{obs}) = \int_0^{z_\mathrm{max}} f_\nu(\nu_\mathrm{obs},z) \frac{\mathrm{d}R}{\mathrm{d}z}(z) \, \mathrm{d}z \, ,
    \end{equation}
    where $dR/dz$ is the differential GW event rate given by each BBH formation channel 
    (see Sections~\ref{sec:ABH} and \ref{sec:PBH} for astrophysical and primordial models, respectively) and $f_\nu$ the energy flux per unit frequency emitted by a source at a luminosity distance $d_\mathrm{L}(z)$ which is related to the comoving distance, $d_c(z)=c/H_0 \int_0^z E(z')^{-1} \mathrm{d}z$ where $E(z)=\sqrt{\Omega_m(1+z)^3+\Omega_\Lambda}$, by the relation $d_\mathrm{L}(z) = (1+z) \, d_\mathrm{c}(z)$.
    Here, the integration limit $z_\mathrm{max}=\frac{\nu_\mathrm{cut}}{\nu_\mathrm{obs}}-1$ is given by the maximal emitting GW frequency $\nu_\mathrm{cut}$. We can express
    \begin{equation}
        f_\mathrm{\nu}(\nu_\mathrm{obs},z) = \frac{1}{4\pi d_\mathrm{L}^2(z)} \frac{\mathrm{d}E}{\mathrm{d}\nu_\mathrm{obs}} = \left . \frac{1}{4\pi d_\mathrm{c}^2(z)} \frac{1}{1+z} \frac{\mathrm{d}E}{\mathrm{d}\nu} \right|_{\nu=\nu_\mathrm{obs}(1+z)},
    \end{equation}
    where we used the coordinate transformation $\mathrm{d} \nu / \mathrm{d} \nu_\mathrm{obs} = (1+z)$ to change from the observer frame to source frame of reference. Here, $\mathrm{d}E/\mathrm{d}\nu$ is the GW energy spectrum emitted by the BBH system evaluated in the source frame. 
    Assuming the BBH systems are in quasi-circular orbits when they reach the $[10^{-4},10^{3}]\,\mathrm{Hz}$ band, we approximate $\mathrm{d}E/\mathrm{d}\nu$ using Eq.~(\ref{eq:dEdnu}) and the waveform approximations for non-precessing spinning BBHs by \citet{2011PhRvL.106x1101A}, as explained in Appendix~\ref{app:GWdEdnu}. Ignoring the precession of spins in the waveforms approximation will not affect our results as the majority of spins in the considered channels are small. In Figure~\ref{fig:dEdnu} we show the GW energy spectrum $\mathrm{d}E/\mathrm{d}\nu$ for BBH systems with varying component masses and $\chi$.

\subsection{Astrophysical binary black hole merger rates}\label{sec:ABH}

    For each astrophysical channel, we define $w_{i,j,k}^\mathrm{intrinsic}$ to be the contribution of a binary $k$ to the intrinsic GW event rate \citep[cf. e.g.,][]{2020A&A...635A..97B,2021arXiv210615841B,2021A&A...647A.153B}.\footnote{For the GC model \citep{2019PhRvD.100d3027R} we take the weights as released by \citet{2021ApJ...910..152Z} where we divide out the coefficient $(1+z_i)^{-1} \left.\frac{\mathrm{d}V_\mathrm{c}}{\mathrm{d}z}\right|_{z_i}$ and multiplied each weight by $\Delta V_\mathrm{c}(z_i)$ to obtain $w_{i,j,k}^\mathrm{intrinsic}$.} This binary is described by component masses $m_{1,k}$ and $m_{2,k}$, and spin vectors ${\bf{a}}_{1,k}$ and ${\bf{a}}_{2,k}$. Each binary is placed at the cosmic time bin $\Delta t_i$ with its redshift of formation $z_{\mathrm{f},i}$ at the center of the bin, and merges at redshift $z_{\mathrm{m},i,k}$ for its corresponding metallicity bin $\Delta Z_j$. Here, the ``intrinsic'' superscript indicates that we assume an infinite detector sensitivity and thus detection probabilities of $p_{\mathrm{det},i,k}=1$, following the notation of Eq.~(D.4) in \citet{2021arXiv210615841B}. Therefore, the BBH event rate observed on Earth for a detector with infinite sensitivity is $N=\int \frac{\mathrm{d}R}{\mathrm{d}z} dz = \sum_{\Delta t_i, \Delta Z_j, k} w_{i,j,k}^\mathrm{intrinsic}$. We can thus calculate the SGWB energy density spectrum of Eq.~(\ref{eq:OmegaGW}), given any arbitrary intrinsic event rate normalisation $N'$, as
    \begin{equation}
        \Omega_\mathrm{GW}(\nu_\mathrm{obs}) = \frac{\nu_\mathrm{obs}}{c^3\rho_c} N' \sum_{\Delta t_i, \Delta Z_j, k} f_{\nu}(\nu_{\mathrm{obs}},z_{\mathrm{m},i,k})\, \tilde{w}_{i,j,k}^\mathrm{intrinsic} \, ,
        \label{eq:OmegaGWastro}
    \end{equation}
    where $\tilde{w}_{i,j,k}^\mathrm{intrinsic}=w_{i,j,k}^\mathrm{intrinsic} / \sum_{\Delta t_i', \Delta Z_j', k'} w_{i',j',k'}^\mathrm{intrinsic}$ is the normalized cosmological weight. The normalisation constants $N'$ are given by the model selection result of \citet{2021arXiv210503349F}. For each considered astrophysical channel we have a median intrinsic event rate value for the Universe observed on Earth of $N'_\mathrm{CE} = 16729.4\,\mathrm{yr}^{-1}$, $N'_\mathrm{SMT} = 905.1\,\mathrm{yr}^{-1}$ and $N'_\mathrm{GC} = 779.6\,\mathrm{yr}^{-1}$, respectively.

    Similar to the BBH event rates, we can calculate the BBH rate density 
    by dividing the normalised cosmological weight contribution of the binary k by the integrated differential comoving volume over $\Delta z_i$ corresponding to the comic time bin $\Delta t_i$, i.e.
    \begin{equation}
        \mathcal{R}(z) = N' \sum_{\Delta Z_j, k} \tilde{w}_{i,j,k}^\mathrm{intrinsic} / \Delta V_\mathrm{c}(z_i) \,\,\, [\mathrm{Gpc}^{-3}\mathrm{yr}^{-1}] \, ,
    \end{equation}
    where $\Delta V_\mathrm{c}(z_i)= \frac{4\pi c}{H_0} \int_{\Delta z_i} d_\mathrm{c}^2(z) / (E(z)(1+z)) \, \mathrm{d}z$ as in Eq.~(D.2) of \citet{2021arXiv210615841B}.

\subsection{Primordial binary black hole merger rate}\label{sec:PBH}

    For the computation of the rate density of binaries in the primordial model, we closely follow the parametrisation of the merger rate as in \cite{2021arXiv210503349F} and Refs. therein. We refer the reader to App.~\ref{app:PBHs} for more details on the predictions of the PBH scenario.
 
    Depending on the initial abundance  $f_\PBH$ and mass function $\psi(m)$, one can estimate the probability of forming binaries in the early Universe due to PBHs decoupling from the Hubble flow. The initial distribution of orbital parameters is defined by the spatial distribution of the surrounding population of PBHs, as well as density perturbations adding an initial torque to the binary system (see e.g. \cite{Ali-Haimoud:2017rtz}). Then, one can compute the differential PBH merger rate density as a function of masses using \citep{DeLuca:2020qqa}
    \begin{align}
    \label{diffaccrate}
    &\frac{ \d^2 \mathcal{R}_\PBH}{\d m^\ii_1 \d m^\ii_2}
     = \frac{1.6 \times 10^6}{{\rm Gpc^{3} \, yr}}
    f_\PBH^{\frac{53}{37}} 
    \eta^{-\frac{34}{37}}(z_\ii)
    \lp \frac{t}{t_0} \rp^{-\frac{34}{37}}  
     \lp \frac{M^\ii}{M_\odot} \rp^{-\frac{32}{37}}  \nonumber \\
    & \times
    S\lp M^\ii, f_\PBH ,\psi  \rp
    \psi(m^\ii_1, z_\ii) \psi (m^\ii_2, z_\ii) 
    \lp \frac{ M(z_\text{\tiny cut-off})}{M (z_\text{\tiny i}) }
    \rp^{9/37} 
    \nonumber \\
    & \times 
    \lp \frac{\eta (z_\text{\tiny cut-off})}{\eta (z_\text{\tiny i}) } \rp^{3/37} 
     \exp\lp \frac{12}{37}\int_{t_\ii} ^{t_\text{\tiny cut-off}} \lp \frac{\dot M(t)}{M(t)} + 2 \frac{\dot \mu(t)}{\mu(t)} \rp \d t 
    \rp 
    ,
    \end{align}
    where $M = m_1 +m_2$, $\mu= m_1 m_2/M$, $\eta = \mu/M$, $t$ is the cosmic time and $t_0$ is the current age of the Universe. 
    Integrating Eq.~\eqref{diffaccrate} in both masses provides the PBH merger rate density as a function of redshift which we adopt in the following. 
    This formula also accounts for the corrective factors introducing the evolution of PBH masses from the initial redshift $z_\ii$ ($\ii$ generally indicates high-$z$ quantities before PBH accretion took place) to the cut-off redshift $z_\text{\tiny cut-off}$ and the shrinking of the binary semi-major axis due to accretion~\citep{DeLuca:2020qqa}. This effect drives the binary evolution up to $z_\text{\tiny cut-off}$, after which the binary evolves through the energy loss  induced by GW emission~\citep{Peters:1963ux}. The suppression factor $S ( M_\text{\tiny tot}, f_\PBH,\psi  ) \equiv S_1 \times S_2$  accounts for both the effect of surrounding DM matter inhomogeneities (not in the form of PBHs) and the disruption of binaries due to interactions with neighbouring PBHs~\citep{Raidal:2018bbj,Vaskonen:2019jpv,Young:2020scc,2021PhRvL.126e1302J,Jedamzik:2020ypm,2020JCAP...11..028D,Tkachev:2020uin,Hutsi:2020sol}.
    In particular, the second piece $S_2$ specifically includes the effect of  disruption of PBH binaries in early sub-structures formed throughout the history of the universe.\footnote{Recent numerical results on PBH clustering confirm the suppression factors we adopt taking into account that the ratio of PBH binaries not entering in clusters is sizeable~\citep{link}.} The two pieces read \citep{Hutsi:2020sol}
    \begin{align}		\label{S1}
    	&S_1 (M, f_\PBH, \psi)  \thickapprox 1.42 \lp \frac{\langle m^2 \rangle/\langle m\rangle^2}{\bar N(y) +C} + \frac{\sigma ^2_\text{\tiny M}}{f^2_\PBH}\rp ^{-21/74} \exp \lp -  \bar N \rp ,
    	\nonumber \\
    	&S_2 (x)  \thickapprox \text{min} \llp 1, 9.6 \cdot 10^{-3} x ^{-0.65} \exp \lp 0.03 \ln^2 x \rp  \rrp, 
    \end{align}
    with $x \equiv  f_\PBH  (t(z)/t_0)^{0.44}$ and the number of neighbouring PBHs being $\bar N \equiv M/\langle m \rangle \times f_\PBH /{ \lp f_\PBH+ \sigma_\text{\tiny M}\rp}$. The constant $C$ appearing in Eq.~\eqref{S1} is defined in Eq.~(A.5) of~\cite{Hutsi:2020sol}.
    Notice that, for $f_\PBH \lesssim 0.003$, one always finds $S_2\simeq 1$, i.e. the suppression of the merger rate due to disruption inside PBH clusters is negligible. This is supported by the results obtained through a cosmological N-body simulation finding that PBHs are essentially isolated for a small enough values of the abundance~\citep{Inman:2019wvr}.
    Therefore, for the small values of $f_\PBH$ adopted in our analysis following \citet{2021arXiv210503349F}, the clustering of PBHs does not play a significant role~\citep{Inman:2019wvr,Vaskonen:2019jpv,2020JCAP...11..028D}.

    In order to compute the SGWB energy density $\Omega_\mathrm{GW}(\nu_\mathrm{obs})$ emitted by the PBH population, we calculate the differential merger rate as a function of redshift as
    \begin{equation}
       \frac{\d R}{\d z} = \int \frac{1}{1+z} \frac{\d V_c (z)}{\d z}  \frac{\d^2 {\cal R}_\PBH }{\d m_1 \d m_2} \d m_1 \d m_2 
    \end{equation}
    and feed this information in Eqs.~\eqref{eq:OmegaGW} and \eqref{eq:Fnu}. Finally, the energy spectrum $\mathrm{d}E/\mathrm{d}\nu$ is computed by integrating over the distribution of PBH binary masses as implied by Eq.~\eqref{diffaccrate}. Also in this case, consistently with the previous section, the PBH model hyper-parameters (i.e. $[M_c, \sigma]$ determining the PBH mass distribution, the abundance $f_\PBH$ and  $z_\text{\tiny cut-off}$ characterising PBH accretion, see more details in App.~\ref{app:PBHs}), are assumed to be given by the population inference result of \citet{2021arXiv210503349F}. In particular, we adopt $ M_c = 34.54 M_\odot$, $\sigma = 0.41$, $f_\PBH = 10^{-3.64}$ and $z_\text{\tiny cut-off} = 23.90$, such that the PBH channel is compatible with explaining around $(1-21)\%$ of the detectable events in the O1/O2/O3a run of LVC, and given the associated astrophysical models considered, the mass gap event GW190521. 
 
\begin{figure*}[h]
\centering
\includegraphics[width=0.75\linewidth]{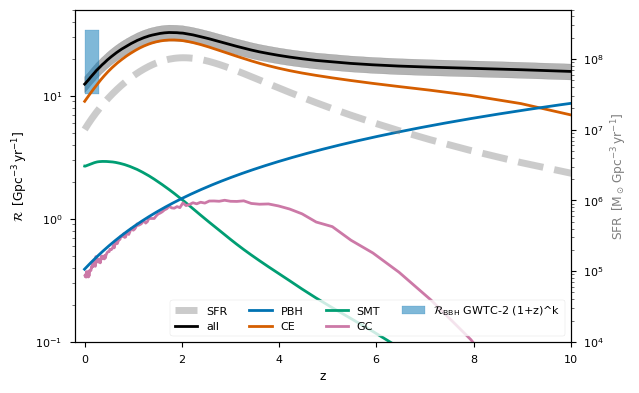} 
\caption{Binary black hole rate density evolution normalised to the model selection results of \citet{2021arXiv210503349F} compared to the local merger rate density inferred from GWTC-2 events assuming a power-law evolution of the merger rate with redshift. We show, in black, the contribution of all channels with relative Poisson error, in gray, computed on the detection rate of the 44 confident BBH events in the O1/O2/O3a observing runs. The model prediction can be directly compared to the local rate constrained by GWTC-2 displayed in blue. With different colors we show the contribution of each channel: common envelope (CE), stable mass transfer (SMT), globular cluster (GC) and primordial black holes (PBH). For comparison, we show the assumed SFR for isolated BBHs which the CE and SMT channels follow, in dashed gray.
}
\label{fig:R_BBH}
\end{figure*}

\subsection{Power-law-integrated sensitivity curves}

    The power-law-integrated (PI) sensitivity curve of a given detector \citep[see][]{2013PhRvD..88l4032T} can be computed once the noise spectral density and the averaged overlap functions are known following the procedure detailed in Appendix~\ref{app:PI}. For the extended LVC network, we assume instrumental noise in different detectors to be uncorrelated. For triangular detectors we take into account the fact that the three nested inteferometers have correlated noise. All PI sensitivity curves are computed using the public code \texttt{schNell} \citep{2020PhRvD.101l4048A}. We choose a (optimistic) SNR threshold of $\rho=2$ to claim a detection, as commonly done in the literature on the subject \citep[see e.g.,][]{2021PhRvD.103d3002P}. Notice that the PI sensitivity curves scale linearly with $\rho$, see Eq.~(\ref{eq:Omega_PI}).   
    
    The current official model for the power spectral density of the LISA noise (for both auto and cross-correlations) is based on the Payload Description Document, and is referenced in the LISA Strain Curves document LISA-LCST- SGS-TN-001.\footnote{We stress that some care has to be taken when comparing different PI sensitivity curves in the literature, as different references assume different arm length for LISA, different observation times and different SNR threshold $\rho$, see e.g. \citet{Petiteau-Note} for an overview. We consider the official configuration with $2.5\,\mathrm{Gm}$ arms and $4\,\mathrm{yr}$ of activity. Our results for the PI agree with Fig. 11 of \cite{2021arXiv210801167B}.}
    
    For ET one might argue that any estimate of correlated noise is quite arbitrary at the moment. We assume that the noise in ET is 20\% correlated between detectors with an arm in common.  This means that magnetic noise lies about a factor 2 in amplitude below other instrument noise. This is a robust assumption at lowest frequencies (< 20Hz). At higher frequencies, if one considers subtraction of magnetic noise, the correlation is expected to be substantially less than 20\%. It is difficult to predict how much one can remove magnetic noise from data with subtraction, but it could even be by another factor 10 in amplitude. A level of correlation of 20\% is therefore quite conservative. As the site for ET has not yet been chosen, we arbitrarily choose a location in Sardinia, close to one of the surveyed sites (see \cite{2020PhRvD.101l4048A} for  the specific coordinates and orientation angle of the triangular network). We use the sensitivity curve of the instrument in the so-called D-configuration \citep{2011CQGra..28i4013H}. The resulting PI sensitivity curve is in quantitative agreement with the one of \cite{2020JCAP...03..050M}.

\section{Results}\label{sec:results}


    We compute the SGWB of astrophysical and primordial BBHs models given the relative event rate contribution determined by the model selection comparison against GWTC-2 events of \citet{2021arXiv210503349F}. For each channel, in Figure~\ref{fig:R_BBH}, we show the BBH merger rate density as a function of redshift as well as their combination. Given our event rate normalisation against the observed 44 events of GWCT-2, the modelled combined local rate density is consistent with the LVC redshift-dependent estimation at $19.3_{-9}^{15.1}\,\mathrm{Gpc}^{-3}\mathrm{yr}^{-1}$ \citep{2021ApJ...913L...7A}\footnote{Notice that the new GWTC-3 rate estimate $17.3-45\,\mathrm{Gpc}^{-3}\mathrm{yr}^{-1}$ at $z= 0.2$ is consistent with the GWTC-2 estimate \citep{2021arXiv211103634T}.}. For comparison, in Figure~\ref{fig:R_BBH}, we also plot the assumed SFR density of the Universe. The rate density redshift evolution of the CE and SMT channels follow the SFR density. The GC BBH rate density does not mimic the star formation history of the host galaxies, instead peaking at $z\in[2,3]$ \citep[see e.g.,][]{2018ApJ...866L...5R}. 
    In contrast to the astrophysical channels, primordial BBHs have a monotonically increasing merger rate density with redshift as~\citep{Ali-Haimoud:2017rtz,Raidal:2018bbj,DeLuca:2020qqa}
    \begin{equation}\label{redevo}
    \mathcal{R}_\mathrm{PBH} (z) \thickapprox \left ( \frac{ t(z)}{t (z=0)} \right)^{-34/37}, 
    \end{equation}
    extending up to redshifts $z={\cal O}(10^3)$.
    Notice that the evolution of the merger rate with time shown in Eq.~\eqref{redevo} is entirely determined by the binary formation mechanism (i.e. how pairs of PBHs decouple from the Hubble flow) before the matter-radiation equality era. 
    Eq.~\eqref{redevo} is, therefore, a robust prediction of the PBH model assuming the standard formation scenario where PBHs are generated with an initial spatial Poisson distribution.
    In our case, given the local normalisation we assume, the PBH contribution grows to overcome the rate density of the astrophysical channels at high redshift ($z>9.5$). 
    In contrast, the BBH merger rate density at low redshifts is dominated by the CE and SMT channels up to $z=2$ where the PBH and GC dominate over SMT. The SMT channel shows a stronger rate suppression over redshift compared to the CE channel as their delay times, i.e. the time between binary formation and BBH merger, are much longer. This occurs because the second MT episode is not as efficient as the CE phase to shrink the BH-Wolf-Rayet binary systems progenitors of the BBHs \citep{2021A&A...647A.153B}.
        

\begin{figure*}[h]
\centering
\includegraphics[width=0.75\linewidth]{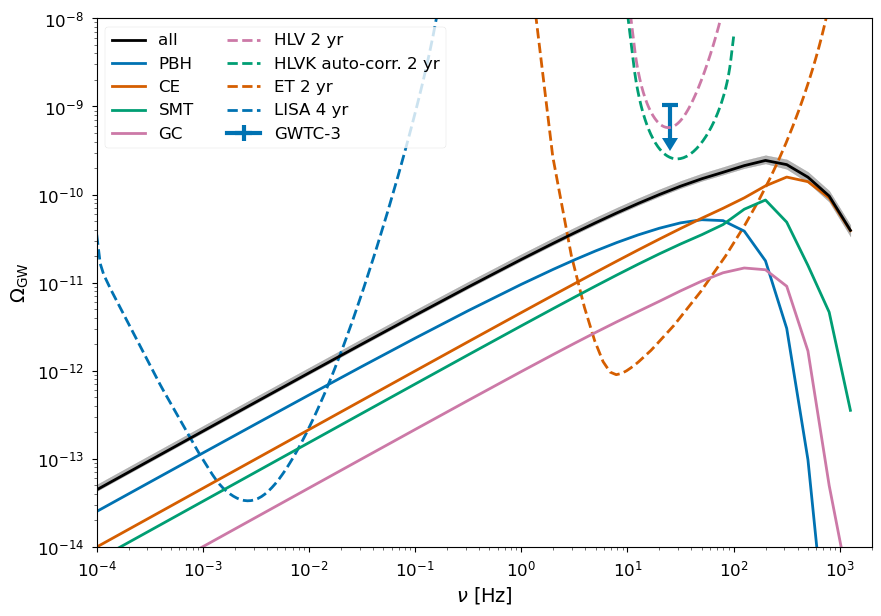} 
\caption{Stochastic gravitational-wave background energy density spectrum of merging astrophysical and primordial BBHs (solid lines). The fiducial model assumes combined BBH event rate normalized against the 44 confident BBH detections of GWTC-2 and branching ratios of each channel inferred by the model selection analysis of \citet{2021arXiv210503349F}. We show with individual lines the partial contribution of each channel: common envelope (orange), stable mass transfer (green), globular cluster (pink) and primordial black holes (blue). The total SGWB is indicated in black where the gray band indicates the Poisson error. The upper constraint to the SGWB from GWTC-3 is indicated with a blue bar marker and an arrow. For comparison, we indicate with dashed lines the PI sensitivity curves of different detectors for corresponding continuous observation time. The detector configurations include LIGO-Virgo at design sensitivity (HLV), the same configuration including KAGRA with auto-correlations (HLVK auto-corr.), Einstein Telescope (ET) and the Laser Interferometer Space Array (LISA). 
}
\label{fig:Omega_GW}
\end{figure*}

    The SGWB energy density spectrum of astrophysical and primordial BBH channels is shown in Figure~\ref{fig:Omega_GW}. We find that, even though the BBH merger rate density is dominated by the CE channel up to redshifts $z\simeq 9.5$, $\Omega_\mathrm{GW}$ is dominated by the PBH channel in the frequency range $\nu\in[10^{-4},400]\,\mathrm{Hz}$. The dominance of the PBH channel over astrophysical channels is explained by two factors. First, contrary to the astrophysical channels, the merger rate density of primordial BBHs is a monotonic increasing function (see Figure~\ref{fig:R_BBH}) peaking at high redshifts of $z \gtrsim 9.5$. Second, given the model inference results of \citet{2021arXiv210503349F}, primordial BBHs are more massive than those produced by the astrophysical channels (see Figure~\ref{fig:observable_dist}) whose coalescence will lead to more energetic GW signals (see Figure~\ref{fig:dEdnu}). The combinations of these two factors leads to the dominance of the PBH channel in the SGWB since $\Omega_\mathrm{GW}$ accounts for the integration over all redshifts. Furthermore, we find that the astrophysical channels dominate $\Omega_\mathrm{GW}$ at $\nu \in [400,1100]\,\mathrm{Hz}$. The suppression of $\Omega_\mathrm{GW}$ at higher frequencies is explained by the fact that the astrophysical channels can produce less massive BBHs compared to our fiducial PBH model (see Figure~\ref{fig:observable_dist}). Given that the inspiral and merger frequency scales as $\nu_\mathrm{merger}\propto M^{-1}$ (cf. Eq.~\ref{eq:nus}, but also see Figure~\ref{fig:dEdnu}) the astrophysical channels contribute to $\Omega_\mathrm{GW}$ above the PBH channel frequency turning point.
    
    Our model predicts\footnote{We report the value of $\Omega_\mathrm{GW}$ at $\nu = 25\,\mathrm{Hz}$ where the current LVC sensitivity has its maximum, see Figure \ref{fig:Omega_GW}.} $\Omega_\mathrm{GW}(\nu = 25\,\mathrm{Hz}) = 1.11_{-0.05}^{+0.16} \times 10^{-10}$, roughly ten times smaller than the current observational upper limits by the LVC of $\Omega_\mathrm{GW}(\nu = 25\,\mathrm{Hz}) \leq 1.04 \times 10^{-9}$ \citep{2021arXiv211103634T}. 
    Notice that this is not a trivial result generated by our normalisation assumption but rather a new  constraint to model selection since these SGWB estimates include the integration over all redshifts whereas the constraints from GW events only probe the models up to the detectors horizons ($z\lesssim1$). 
    Further, we can predict whether our BBH SGWB estimate will be detected by future GW observing runs. To this end, in Figure~\ref{fig:Omega_GW}, we show the PI sensitivity curves for the ground base detectors LIGO-Virgo (HLV), LIGO-Virgo-KAGRA (HLVK) with auto-correlations and ET accounting for $2\,\mathrm{yr}$ of integrated data collection at their design sensitivities as well as the space-based detector LISA for the nominal $4\,\mathrm{yr}$ mission. Given our model assumptions, we find that the SGWB of merging BBH lies below the current generation of GW detectors. However, we find that LISA and the third-generation (3G) GW detectors such as ET will be able to detect the BBH SGWB.


\section{Discussion}\label{sec:discussion}

    
    

    Compared to the analysis of \citet{2021ApJ...910..152Z}, in \citet{2021arXiv210503349F} the CHE 
    channel was excluded, while the inclusion of the NSC channel was not found to lead to a larger Bayesian evidence, still reaching similar conclusions regarding the relative contributions of CE, SMT and GC channels. This is due to the CHE and NSC models contributing $< 8\%$ and $< 13\%$, respectively, to the underlying merging BBH distribution in \citet{2021ApJ...910..152Z} at 95\% credibility. More precisely, the CHE channel mostly leads to highly spinning and massive BHs which are favoured due to GW detector selection effects and a smaller contribution to the underlying BBH distribution is required to explain the events in GWTC-2. A similar argument can be made for the NSC channel where in contrast to CHE channel BH spins are smaller. Given the small predicted contribution of these two channels to the intrinsic merger rate distribution we expect a subdominant contribution to $\Omega_\mathrm{GW}$, much smaller than the GC contribution in Figure~\ref{fig:Omega_GW}, and hence, will not affect our results.
    
    
    Our analysis neglected the contribution from the population of non-merging binaries, which previous studies on isolated binary evolution \citep{2021PhRvD.103d3002P} have found to be negligible compared to the merging population in the frequency bands we consider. Furthermore, we also did not include eccentric corrections to the GW energy spectrum and assumed that the BBHs will reach the LISA or ground based detector sensitive frequencies with quasi-circular orbits. \cite{2021PhRvD.103d3002P,2021arXiv211201119P} showed that this is the case for isolated binary evolution, while a number of semi-analytic and numerical studies~\citep[][]{2016ApJ...830L..18B,2017ApJ...840L..14S, 2018PhRvL.120o1101R,2018MNRAS.481.5445S,2019ApJ...871...91Z,2021arXiv210609042Z} have shown that for dynamically formed systems a sizable fraction of the BBH population can retain appreciable eccentricity when they enter the LIGO or LISA bands. However, since the GC channel shows a subdominant contribution to the total $\Omega_\mathrm{GW}$ at one part in 20 at $\nu = 3\,\mathrm{mHz}$, any boost to the GW background from this channel will not change our conclusion significantly. 
    
    In this work we presented our SGWB analysis as an additional constraint to multi-channel Bayesian model selection where $\Omega_\mathrm{GW}$ is computed a posteriori to be much smaller than the observational constraint. Alternatively, if we had a more stringent constraint to $\Omega_\mathrm{GW}$, one could include the SGWB constraint in the inference of such model selection frameworks, providing the analysis with a direct discriminatory power to models overpredicting $\Omega_\mathrm{GW}$. Because in our analysis the SGWB is found to be much smaller than the LVC observational constraint, the result of our model selection would remain the same. On a similar end, given a phenomenological model, \cite{2020ApJ...896L..32C} recently used the SGWB and informations about individually resolvable events from O1 and O2 to place joint constraints on the BBH merger rate density peak and slope for low redshifts. Notice that the two analysis are complementary as the one presented here simultaneously constrains the potential contribution of each channel to the total BBH merger rate density while \cite{2020ApJ...896L..32C} puts a direct constraints on the overall joint merger rate density. 
    

    A future detection of the SGWB would also possibly help in distinguishing between astrophysical and primordial
    channels through the SGWB anisotropies \citep{2021arXiv210701935W}. Another example for distinguishing between channels is that in the case the GC channel dominates the SGWB, one would expect to be able to identify a cusp in $\Omega_\mathrm{GW}$ in LISA frequency band \citep{2018MNRAS.481.4775D}. Other possible ways to discriminate among different channels is the measurement of the  merger bias at 3G detectors through the study of the cross-correlation with the large scale structure \citep{2017PhRvD..96j3019C, 2018PhRvL.120w1101C,2018JCAP...09..039S,2020arXiv200801082M,2020MNRAS.491.4690M,2020PhRvR...2b3314C,2021MNRAS.500.1666Y}, the study of the time evolution of the high redshift merger rates \citep{2021JCAP...05..003D,2021arXiv210807276N}, and the reconstruction of the spectral shape pre-merger via small band searches. 
    
    
\section{Conclusions}\label{sec:conclusions}
    
    With the aim of further constraining multi-channel BBH Bayesian model selection, as a working example in this work, we computed the SGWB of an astrophysical and primordial BBH population resulting from a model selection comparison with GWTC-2. Because our study did not include all potentially prominent channels leading to merging BBHs or span all model uncertainties (cf. Section~\ref{sec:intro}), the results of the analysis need to be carefully interpreted. Rather than being a definitive answer to which channel dominates the SGWB, our analysis is intended to provide a complementary tool to model selection that can be used to further constrain the contribution of particular formation channels. In principle, it provides an additional constraint to model selection as the SGWB probes the theoretical prediction over all redshifts. This is in contrast to model selection analyses that focus on the resolved population, which have discriminating power only up to the GW detector horizons, currently at $z\lesssim1$ for LIGO--Virgo. Our methodology can therefore be extended to all future model selection studies aiming to unravel the origin of merging BBHs independently from which channel dominates the BBH merger rate density.
    
    To mitigate the large model uncertainties on the merger rate density estimation of each channel which directly correlate to $\Omega_\mathrm{GW}$, we have normalized the event rate to the 44 confident detections of GWTC-2 and the relative contribution of each channel to the branching ratio results in the model selection analysis of \cite{2021arXiv210503349F}. 
    We predict a SGWB of $\Omega_\mathrm{GW}(\nu=25\,\mathrm{Hz})=1.11^{+0.16}_{-0.05}\times 10^{-10}$ which lies below the current observational upper limits published by the LVC. Moreover, we find that such background will be accessible only to 3G GW observatories such as ET and the space-based detector LISA.
    
    
    
    Finally, with 3G Earth-based detectors, a catalogue of individual events and background mapping provide complementary information on the underlying source population. Combining the two approaches 
    can be useful to gain information on a high redshift population of sources which cannot be detected and studied individually, even with large-horizon instruments. 

\begin{acknowledgements}
    The authors would like to thank Valerio De Luca for useful discussions on early stages of the project.
    This work was supported by the Swiss National Science Foundation Professorship grant (project number PP00P2 176868).
    G.F. and  A.R. are supported by the Swiss National Science Foundation (SNSF), project {\sl The Non-Gaussian Universe and Cosmological Symmetries}, project number: 200020-178787. The work of G.C. is supported by SNSF Ambizione grant - {\sl Gravitational waves propagation in the clustered universe}.
    M.Z. is supported by NASA through the NASA Hubble Fellowship grant HST-HF2-51474.001-A awarded by the Space Telescope Science Institute, which is operated by the Association of Universities for Research in Astronomy, Inc., for NASA, under contract NAS5-26555. 
\end{acknowledgements}

\bibliography{aanda}

\appendix

\section{Binary black hole models}\label{app:BBH_models}

\subsection{Astrophysical BBH models}

\subsubsection{Isolated binary evolution models}

We consider the SMT and CE channels for the formation of merging BBH through isolated binary evolution. In both cases a typical binary evolves first through a stable MT episode which is caused by the primary star evolving faster and expanding first to overfill the Roche-lobe when the star leaves the main sequence to become a red supergiant. Stripped from its envelope, the primary eventually collapse to form a BH. Similarly, when the companion star leaves its main sequence, the secondary will expand to overfill the Roche-lobe leading to the second mass transfer episode which can be either be stable (SMT channel) or unstable (CE channel). The latter case leads to a CE phase where the envelope of the secondary engulf the BH companion. If the binary survives the MT episode, namely it avoids merging, a BH-Wolf-Rayet systems is formed. Compared to the SMT channel, the CE channel leads to smaller orbital separations post MT \citep{2021A&A...647A.153B}. For BH-Wolf-Rayet orbital periods smaller than 1 day tidal interactions from the BH onto the companion can lead to the spin-up of the Wolf-Rayet star which subsequently leads to the formation of a highly spinning second-born BH \citep{2018A&A...616A..28Q,2020A&A...635A..97B,2021A&A...647A.153B,2021RNAAS...5..127B}.

In this work we adopt the CE and SMT models of \citet{2021A&A...647A.153B} which used \texttt{POSYDON}\footnote{\url{https://posydon.org}} (Fragos et al. 2021, in prep.) to combine the rapid parametric population synthesis code \texttt{COSMIC} \citep{2020ApJ...898...71B} with detailed \texttt{MESA} \citep{2011ApJS..192....3P,2013ApJS..208....4P, 2015ApJS..220...15P,2018ApJS..234...34P,2019ApJS..243...10P} binary evolution simulations. \texttt{COSMIC} was used to rapidly evolve binaries from zero-age main sequence to post second MT while \texttt{MESA} to target the BH-Wolf-Rayet evolutionary phase leading to the tidal spin-up of the secondary. The spin of the first-born BH is assumed to be zero, a direct consequence of the assumed efficient angular transport \citep{2015ApJ...800...17F,2018A&A...616A..28Q,2019ApJ...881L...1F} which finds support in asteroseismic measurements \citep{2014MNRAS.444..102K,2015A&A...580A..96D,2018A&A...616A..24G}, observations of white dwarfs spins \citep{2005A&A...444..565B} and recent gravitational-wave observations \citep{2021ApJ...910..152Z}. Moreover we assume Eddington-limited accretion efficiency onto compact objects, which leads to negligible mass accretion onto the first-born BH and prevent any mass-accretion spin-up in the SMT channel \citep{1974ApJ...191..507T}. For a detailed description of all the model parameters we point the reader to \citet{2021A&A...647A.153B}.

\subsubsection{Dynamical formation models in GC}

In addition to isolated evolution channels, we also consider the astrophysical formation channel of BBH mergers that are synthesized in dense stellar environments. 
In particular, we use the set of globular cluster models from \cite{2019PhRvD.100d3027R}, which are simulated using the Henon-style cluster Monte Carlo code \texttt{CMC}~\citep{1971Ap&SS..13..284H,1971Ap&SS..14..151H,2000ApJ...540..969J,2013ApJS..204...15P}. 
These cluster models span a range of initial masses, half-mass radii and metallicities and show present-day consistency with observational properties of globular clusters in the Milky Way and nearby galaxies. 
We use the models from \cite{2019PhRvD.100d3027R} in which black holes are born with near-zero spin, as spin has only a minor effect on the GW energy spectrum and this model is preferred by the data~\citep{2021ApJ...910..152Z,2021arXiv210503349F}. 
Cluster formation, and therefore the redshift distribution of mergers, follow the prescriptions in \cite{2018ApJ...866L...5R}, and the cluster population is weighted based on the metallicity distribution of Milky Way globular clusters.

\subsection{Primordial BBH model}\label{app:PBHs}

The formation of PBHs occurs from the collapse of large overdensities in the primordial Universe \citep[see][for a recent review]{2021JPhG...48d3001G}. 
The formation of a PBH of mass $m$ takes place deep in the radiation-dominated era at a typical redshift $z_\ii \simeq 2 \cdot 10^{11} (m/M_\odot)^{-1/2}$.
The distribution of masses is determined by the characteristic size and statistical properties of the density perturbations, corresponding to curvature perturbations 
generated during the inflationary epoch. 
As typically done in the literature, we assume a  model-independent parametrization of the mass function at formation redshift $z_\ii$ of the form
\begin{equation}\label{mass dist}
\psi(m,z_\ii) = \frac{1}{\sqrt{2\pi}\sigma  m} \exp \lp -\frac{\log ^2 (m/M_c)}{2 \sigma^2} \rp,
\end{equation}
in terms of its width $\sigma$ and reference mass scale $M_c$ (not to be confused with the chirp mass denoted here with ${\cal M}$). 
This mass function describes a PBH population 
resulting from a symmetric peak in the curvature spectrum and recovers a wide variety of models \citep{Dolgov:1992pu,Green:2016xgy,Carr:2017jsz}.\footnote{In the literature, other PBH mass functions were also considered~\citep[e.g.][]{2017PhRvD..95h3508K,2018JCAP...01..004B,2020PhRvD.102l3524H,2020arXiv200903204G}. In this work, we follow
\cite{2021arXiv210503349F} and adopt Eq.~\eqref{mass dist}.} 

As extreme perturbations tend to have nearly-spherical shape~\citep{bbks} and the collapse takes place in a radiation-dominated Universe, the initial adimensional Kerr parameter ${\bf{a}}$ is expected to be below the percent level \citep{DeLuca:2019buf,Mirbabayi:2019uph}. 
However, a non-zero spin can be acquired by PBHs forming binaries through an efficient phase of accretion \citep{DeLuca:2020qqa,DeLuca:2020bjf} prior the reionization epoch. 
Therefore, a defining characteristics of the PBH model is the expected correlation between large values of binary total masses and large values of spins of their PBH constituents. 
Also, the spin directions of PBHs in binaries are independent and they randomly distribute on the sphere. 
PBH accretion is still affected by large uncertainties, in particular coming from the impact of feedback effects \citep{Ricotti:2007jk,Ali-Haimoud:2017rtz}, structure formation \citep{2020JCAP...07..022H,raidalsm} and early X-ray pre-heating \citep{Oh:2003pm}.
Therefore, an additional hyper-parameter, the cut-off redshift $z_\co \in [10,30]$, was introduced by \cite{DeLuca:2020bjf}  accounting for this accretion model uncertainties.
For each value of $z_\co$ there exists a one-to-one correspondence between the initial and final masses which can be computed according to the accretion model described in details in Refs.~\citep{Ricotti:2007jk,Ricotti:2007au,DeLuca:2020qqa,DeLuca:2020bjf}. We highlight, for clarity, that a lower cut-off is associated to stronger accretion and vice-versa. Values above $z_\co \simeq 30$ correspond to negligible accretion in the mass range of interest for the LVC observations.

In the absence of primordial non-Gaussianities, the PBH locations in space at the formation epoch follow a Poisson distribution~\citep{Ali-Haimoud:2018dau,Desjacques:2018wuu,Ballesteros:2018swv,MoradinezhadDizgah:2019wjf}.
This feature, describing the spatial distribution of PBHs at formation in the stadard scenario, is used to compute the properties of the population of PBH binaries formed at high redshift and contributing to the merger rate described in Sec.~\ref{sec:PBH}.
Notice, finally, that that the dominant PBH merger rate comes from PBH binaries assembled via gravitational decoupling from the Hubble flow before matter-radiation equality \citep{Nakamura:1997sm,Ioka:1998nz}
and, as shown for example in \cite{Ali-Haimoud:2017rtz}, the merger rate of binaries formed through dynamical capture in the present day halos is subdominant and, therefore, neglected in our computations.

\section{Gravitational-wave energy spectrum}\label{app:GWdEdnu}

We approximate the gravitational-wave energy spectrum, $\frac{\mathrm{d}E}{\mathrm{d}\nu}$, of a coalescing BBH using the phenomenological templates models of \citet{2011PhRvL.106x1101A} which are obtained using frequency domain matching of post-Newtonian inspiral waveforms with coalescence waveforms from numerical simulations. These models approximate the inspiral, merger and ringdown waveforms in the frequency Fourier space domain, $\nu$, for BBHs with component masses, $m_1$ and $m_2$, and non-precessing spins, $\bf{a}_1$ and $\bf{a}_2$. 
Assuming circular orbits, we have \citep{2011ApJ...739...86Z}
\begin{equation}
    \frac{\mathrm{d}E}{\mathrm{d}\nu} = \frac{(G \pi)^{2/3} \mathcal{M}^{5/3}}{3} 
    \begin{cases}
     \nu^{-1/3} f_1^2  & \nu < \nu_\mathrm{merger}\\
     \omega_1 \nu^{2/3} f_2^2 & \nu_\mathrm{merger} \leq \nu < \nu_\mathrm{ringdown} \\
     \omega_2 f_3^2 & \nu_\mathrm{ringdown} \leq \nu < \nu_\mathrm{cut} 
    \end{cases}
    \label{eq:dEdnu}
\end{equation}
where in the above expression
\begin{equation}
\begin{split}
    f_1 & \equiv f_1(\nu, M,\eta,\chi) = 1+\alpha_2\nu'^2 + \alpha_3 \nu'^3 \, , \\  
    f_2 & \equiv f_2(\nu, M,\eta,\chi) = 1 + \varepsilon_1 \nu' + \varepsilon_2 \nu'^2 \, ,\\
    f_3 & \equiv f(\nu, \nu_\mathrm{ringdown}, M, \chi) = \frac{\nu}{1+\left(\frac{2(\nu-\nu_\mathrm{ringdown})}{\sigma}\right)^2} \, ,
\end{split}
\end{equation}
with $\nu' = (\pi MG \nu / c^3)^{1/3}$ and
\begin{equation}
    \begin{split}
    & \alpha_2 = -\frac{323}{224}+\frac{451}{168}\eta \,, \\ 
    & \alpha_3  = \left( \frac{27}{8} - \frac{11}{6}\eta \right)\chi \, , \\
    & \varepsilon_1 = 1.4547 \chi - 1.8897 \,,\\
    & \varepsilon_2 = - 1.8153 \chi + 1.6557 \, .
    \end{split}
\end{equation}
Here, the merge, ringdown and cut frequencies, as well as $\sigma$, are approximated by
\begin{equation}
\begin{split}
    &\nu_\mathrm{merger} = \frac{c^3}{\pi M G}  \left(1 - 4.455(1-\chi)^{0.217} + 3.521 (1-\chi)^{0.26} + \mu_\mathrm{merger} \right) , \\
    &\nu_\mathrm{ringdown} = \frac{c^3}{\pi M G}  \left(\left(1-0.63(1-\chi)^{0.3}\right)/2 + \mu_\mathrm{ringdown}\right) ,\\
    &\nu_\mathrm{cut} = \frac{c^3}{\pi M G}  \left(0.3236 + 0.04894\chi + 0.01346 \chi^2  + \mu_\mathrm{cut}\right) , \\
    & \sigma = \frac{c^3}{\pi M G} \left( \left(1-0.63(1-\chi)^{0.3}\right)(1-\chi)^{0.45}/4 + \mu_\mathrm{\sigma} \right),
\end{split}
\label{eq:nus}
\end{equation}
where $\mu^{(ij)}_k$ with $k \in [\mathrm{merger},\mathrm{ringdown},\mathrm{cut},\sigma]$ are computed as
\begin{equation}
    \mu_k \equiv \mu_k(\eta,\chi) = \sum_{i=1}^{3} \sum_{j=0}^{\min(3-i,2)} y_k^{(ij)} \eta^i \chi^j \, ,
\end{equation}
with $y_k$ coefficents given in Table~1 of \citet{2011PhRvL.106x1101A}.
Finally, $\omega_1$ and $\omega_2$ are normalisation constants that guarantee continuity at $\nu_\mathrm{merger}$ and $\nu_\mathrm{ringdown}$, respectively,  
\begin{equation}
 \begin{split}
    & \omega_1 = \nu_\mathrm{merger}^{-1}f_1^2(\nu_\mathrm{merger}, M, \eta, \chi)/f_2^2(\nu_\mathrm{merger}, M, \eta, \chi) \, , \\
    & \omega_2 = \omega_1 \nu_\mathrm{ringdown}^{-4/3} f_2^2(\nu_\mathrm{ringdown}, M, \eta, \chi)  \, .
 \end{split}
\end{equation}

\section{Power-law-integrated sensitivity curve}\label{app:PI}

Let us assume to have a network of $N$ detectors. 
The antenna pattern for a detector pair $AB$ is given by 
      \begin{equation}\label{eq:antenna}
        \mathcal{A}_{AB}({\bf{n}},\nu)=\gamma_{AB}({\bf{n}})\,e^{-i2\pi \nu{\bf{n}}\cdot{\bf b}_{AB}}.
      \end{equation}
These functions are already defined in \texttt{schNell} \citep{2020PhRvD.101l4048A} for different detector classes. 
We define the average antenna patter function, often referred to in the literature as average overlap function (i.e. the exponential factor in Eq.~(\ref{eq:antenna}) is often included in the definition of $\gamma_{AB}$) as
\begin{equation}
\bar{\gamma}_{AB}(\nu)=\int \frac{\mathcal{A}_{AB}(\nu,{\bf{n}})}{4\pi} \mathrm{d}^2{\bf{n}}\,.
\end{equation}
We start from the effective noise power spectral density. In full generality this is given by
\begin{equation}\label{Sf}
S_{\text{eff}}=\left[\sum_{ABCD} (N_{\nu}^{-1})^{AB}\bar{\gamma}_{BC}(\nu)(N_{\nu}^{-1})^{CD}\bar{\gamma}_{DA}(\nu)\right]^{-1/2}\,,
\end{equation}
for the case of uncorrelated detectors 
\begin{equation}
(N^{-1}_{\nu})^{AB}=\frac{\delta_{AB}}{N_{\nu}^A}\,,
\end{equation}
where $N_{\nu}^A$ is the noise power spectral density (PSD). In this case Eq.~(\ref{Sf}) simplifies to
\begin{equation}
S_{\text{eff}}=\left[\sum_{A=1}^N \sum_{B>A}^N \frac{\bar{\gamma}_{AB}}{N_{\nu}^A N_{\nu}^B}\right]^{-1/2}\,.
\end{equation}
However, for triangular detectors such as LISA and ET one needs to know both the noise variance (auto-correlation) $N_{\nu}^{11}$, and the cross-detector covariance $N_{\nu}^{12}$. 
We convert Eq.~(\ref{Sf}) to energy density units using 
\begin{equation}
S_{\text{eff}}(\nu)=\frac{3 H_0^2}{2 \pi^2}\frac{\Omega_{\text{eff}}(\nu)}{\nu^3}\,.
\end{equation}
For a set of power low indices $\beta$, we compute the value of the amplitude $\Omega_{\beta}$ such that the integrated signal to noise ratio $\rho$ has some fixed value, here we assume $\rho=2$. This is given by
\begin{equation}
\Omega_{\beta}=\frac{\rho}{\sqrt{2 T}}\left[\int_{\nu_{\text{min}}}^{\nu_{\text{max}}} \frac{(\nu/\nu_{\text{ref}})^{2\beta}}{\Omega_{\text{eff}}^2(\nu)}\right]^{-1/2} \mathrm{d} \nu\,.
\end{equation}
Note that the choice of $\nu_{\text{ref}}$ is arbitrary and will not
affect the sensitivity curve.
For each pair of values $(\beta, \Omega_{\beta})$, we compute $\Omega_{\text{GW}}=\Omega_{\beta}(\nu/\nu_{\text{ref}})^{\beta}$. The envelope is the power low integrated sensitivity curve. Formally it is given by
\begin{equation}
\Omega_{\text{PI}}(\nu)=\text{max}_{\beta}\left[\Omega_{\beta}\left(\frac{\nu}{\nu_{\text{ref}}}\right)^{\beta}\right]\,.
\label{eq:Omega_PI}
\end{equation}
Any line (on a log-log plot) that is tangent to the power-law integrated sensitivity curve corresponds to a gravitational-wave background power-law
spectrum with an integrated signal-to-noise ratio $\rho=2$.
This implies that if the curve for a predicted background lies everywhere below the sensitivity curve, then $\rho<2$ for
such a background.

\end{document}